\documentstyle[preprint,eqsecnum,aps]{revtex}
\begin{document}
\def\om{\omega}
\def\P{\Phi}
\def\p{\Phi}
\def\phi{\Phi}
\def\square{\kern1pt\vbox{\hrule height 1.2pt\hbox{\vrule width 1.2pt\hskip 3pt
   \vbox{\vskip 6pt}\hskip 3pt\vrule width 0.6pt}\hrule height 0.6pt}\kern1pt}

\def\lta{\mathrel{\spose{\lower 3pt\hbox{$\mathchar"218$}}
     \raise 2.0pt\hbox{$\mathchar"13C$}}}
\def\gta{\mathrel{\spose{\lower 3pt\hbox{$\mathchar"218$}}
     \raise 2.0pt\hbox{$\mathchar"13E$}}}
\def\spose#1{\hbox to 0pt{#1\hss}}

\draft
\preprint{CITA-94-18}
\title{Gravity-Driven Acceleration of the Cosmic Expansion }
\author{Janna Levin }
\address{ Canadian Institute for Theoretical Astrophysics}
\address{Mc Lennan Labs, 60 St. George Street, Toronto, ON}
\maketitle
\begin{abstract}

It is shown here that a dynamical Planck mass
can drive the scale factor of the universe to accelerate.
The negative pressure which drives
the cosmic acceleration is identified with the unusual
kinetic energy density of the Planck field.
No potential nor cosmological constant is
required. This suggests a purely gravity driven, kinetic
inflation.  Although the possibility is not ruled out,
the burst of acceleration is often too weak
to address the initial
condition problems of cosmology.
To illustrate
the kinetic acceleration, three different
cosmologies are presented.  One such example,
that of a bouncing universe, demonstrates
the additional feature of being nonsingular.
The acceleration is also considered in the conformally related
Einstein frame in which the Planck mass is constant.

\vskip10pt

98.80.Hw, 98.80.Cq,04.50.+h

\bigskip

\centerline{submitted May 18, 1994}

\end{abstract}

\narrowtext
\vfill\eject

\section{Introduction}

In Einstein's theory of general relativity, the gravitational
constant is proposed to be a universal constant of nature,
$G=1/M_o^2$ where $M_o=10^{19}$ GeV is the constant Planck mass.
A simple extension of Einstein gravity elevates the Planck
mass from a fundamental constant to a dynamical variable.
The strength of gravity is then allowed to evolve as the
universe evolves.
Scalar-tensor theories proposed by Jordan \cite{Jo} and
Brans and Dicke \cite{Brans}
and generalized by Bergmann \cite{Bergmann}
and by Wagoner \cite{Wagoner}
incorporate such a modification of general relativity.
A revival of Jordan-Brans-Dicke (JBD) gravity has occured in the
literature recently.
There has been considerable interest in various JBD cosmologies and their
phenomenological uses from inflationary to stringy cosmologies.

Here the early behavior of a universe empty except for the background
dynamical
Planck mass is studied. No potential
is included. Remarkably, this investigation
reveals the Planck field can accelerate the
expansion of the universe.
A negative pressure is always needed to
accelerate the cosmic expansion.
For ordinary matter,
the pressure associated with kinetic energy
is positive.  By contrast,
the nonminimal coupling of the Planck
field to gravity allows for an unusual negative pressure associated with the
kinetic energy of the field.
It is worth stressing
that there is no  potential\footnote{Although
the nonminimal interaction of the Planck mass
with gravity can in some sense be considered
a potential, the energy density of the interaction
is a function of derivatives
of the Planck mass,  and in that sense is kinetic.
It is for this reason that the acceleration is dubbed kinetic
driven
and not potential driven.}
nor cosmological constant.
This possibility was first pointed out in
in references \cite{un}.

The source of the negative pressure in scalar-tensor
theories of gravity can be understood
heuristically.  Loosley speaking, a pressure measures the negative of
the change in energy  with volume.
If, as the universe expands,
the energy contained within a unit volume decreases,
the pressure is positive.  Usually the energy does decrease
as it takes work to power the expansion.
For instance, a radiation bath redshifts and cools as it does
work on the spacetime, so that the energy
drops as the universe expands, $E_{\rm rad}\propto
\rho_{\rm rad}a^3\propto 1/a\sim 1/V^{1/3}$
where $a$ is the scale factor.
As the energy decreases,  the Hubble expansion slows.
In other words, the scale factor decelerates.
If instead the energy contained within a unit
volume  increases, then the pressure is negative.
As the energy increases, the Hubble expansion quickens;
that is, the universe accelerates.
In standard inflation for instance
\cite{alan}, the energy density is
constant,  $\rho\propto \Lambda$.  As the scale factor grows, the
energy contained within a unit volume grows, $\rho a^3 \propto \Lambda
a^3\propto \Lambda V$, while
$p\propto -\Lambda$ is negative.  The expansion gets ever
faster.

Ordinarily kinetic energy has an associated positive
pressure.
This is not so in general scalar-tensor theories.
The scalar-tensor theory can be defined entirely in
terms of the kinetic coupling parameter $\om(m_{pl})$,
where $m_{pl}$ is the dynamical Planck mass
(see (\ref{una})).
If $\om$ is a positive constant, as is assumed in the original Brans-Dicke
theory, then the pressure associated with the field
is positive.  However, if
$\om$ grows with  the Planck mass so that it satisfies a bound
defined below, then the
kinetic energy in the Planck field can grow  as the universe
expands.
There is therefore a negative pressure and a
corresponding acceleration of the scale factor.
No potential nor cosmological constant is required.
The acceleration is weak since there is a competition between
the $\om$ effect and the redshifting
in the kinetic energy.\footnote{There is one additional possibility.  If
$\om$ is {\it constant but negative} and the    Planck mass
drops, then there exists a branch of the solutions which
leads to an accelerated expansion, as noted in reference \cite{un}.
The kinetic energy grows as the Planck mass drops and the
pressure is negative.
This regime was recently considered, independently, in reference
\cite{Ram}.  Specifically, the authors of
\cite{Ram} considered $\om=-1$ in  a string-dilaton
cosmology.
They also found an accelerated expansion.}

Since an acceleration is one of the key
ingredients used in inflation, this feature is provocative.
It is not attempted in this paper to build a model of gravity
driven inflation.  In fact success in such an attempt may
be unlikely (see \S \ref{kin}).
Instead a nominal condition for the theory
to be pertinent for the causal physics of inflation
is investigated.
Although the universe may accelerate when
$\om(t)>0$, it is demonstrated that
the acceleration can be relevant for
inflation only in theories for which
$\om$ is allowed to drop below zero
[and then, only when the Planck
mass drops as the universe evolves].
As demonstrated in \S \ref{aqua}, the energy density in
a Friedman-Robertson-Walker universe is
positive as long as $\om\ge -3/2$.  The range
$-3/2\le \om<0$
is therefore not ruled out by the weak energy condition.
The positivity of the energy density  can be verified
by looking in the conformally related Einstein frame as
is done in \S \ref{conform}.  These restrictions apply only
for the simplest single scalar model.
Chaotic alternatives have still to be explored.
Further difficulties in building a sound kinetic inflationary
model are discussed briefly in \S \ref{kin} and more
fully in \cite{prep}.

Independent of inflation, if the universe accelerates,
an initial singularity may be circumvented.
Loosely speaking,
if the universe accelerates at its
inception then the Hubble expansion is not infinite initially.
Consequently, as one looks back in time to the first moment, the
scale factor is not forced to zero and
may approach a finite value.
One example, expounded below, exhibits such
nonsingular behavior.

\section{Three cosmic accelerations}

The simplified situation
of a universe empty, except for the background Planck field
is studied in this paper.  The metric is assumed to be a flat
Friedman-Robertson-Walker (FRW) metric.
The treatment is purely classical.
Under these simplified
conditions it is shown that an epoch of cosmic acceleration
can in fact ensue.
Three examples are given in which the acceleration is manifest.

The first example is of a bouncing universe.
The universe begins infinitely large and contracts.
It collapses down to finite size and then
bounces into an accelerating phase.
Alternatively, the universe could be created
at the moment of the bounce in a nonsingular
beginning.  If this initial condition
is imposed, the
universe is created cold and empty
except for the background Planck field.
The energy density is zero initially.
It thus costs no energy to create such a universe.
The Planck field begins to move and as it does so the
universe begins to expand.
The changing structure
of gravity accelerates the expansion
from zero.
Eventually, the change in the parameter $\om$ becomes
more temperate and the expansion decelerates.

The second example is constructed so that
$\om$ never drops below zero, in order to assure the reader
that such a model can be built with an accelerating phase.
The universe begins singular in this example and decelerates
as the universe evolves. As the Planck field enters a range
of values, a weak burst of acceleration ensues.  It turns
off naturally and the universe decelerates from that moment
until   the end of time.

The third example is the  only one in this paper
which satisfies the nominal requirement
relevant for the causal
physics of inflation.
It is not intended as a true model of inflation, only as
an indication of the properties
a kinetic inflationary model would have.
The possible pitfalls and merits of a serious attempt
at building such a
model are discussed in that section.

For completeness, the kinetic acceleration is studied
in the conformally related Einstein frame.
In the Einstein frame, the Planck field is constant.
An Einstein observer believes the
universe's expansion decelerates.
However the distance
between two points in space, relative to ruler lengths,
agrees with the original JBD frame.
Said another way,
the acceleration is attributed
to the relative
rate of change of
the length of an observer's rulers.

\bigskip

Before proceeding, some of the recent cosmological settings in which
scalar-tensor theories of
gravity have been considered are listed.
Solutions describing the universe have been found
for  JBD theories
in vaccum  \cite{Barrow}, in a radiation-dominated
era \cite{Barrow}, \cite{Lars},\cite{us1},\cite{us2},
as well as in inflationary
models such as (hyper)extended inflation \cite{Stein}
or induced gravity inflation \cite{induce}.
Recently the suggestion has been made that for
certain theories, the presence of nonrelativistic matter
can serve to attract scalar-tensor gravity toward general relativity
\cite{Damour}.
This idea was extended in reference \cite{Polyakov} to low energy
string theories where the dilaton plays the role of the Planck mass.
It is noteworthy that
the first example studied happens, quite coincidentally, to be of
the form used in  \cite{Damour} to illustrate that general
relativity is an attractor
of some scalar-tensor models.

\section{Acceleration from Scalar-Tensor Gravity}
\label{aqua}

For simplicity, in this paper it is assumed
that the universe is created
initially devoid of all contributions to the energy-momentum tensor from
ordinary matter.
The universe studied is therefore empty, except for the
kinetic energy density of the background Planck field.
Beginning with the action for general scalar-tensor theories,
a bound on the changing structure of gravity can be found
such that the cosmic expansion is accelerated.

The gravitational  action for a general Jordan-Brans-Dicke
theory  is
        \begin{equation}{A[g_{\mu\nu},\Phi]=
        \int d^4 x\sqrt{-g}\left [ {\Phi\over 16 \pi}{\cal R}-{\om(\Phi)\over
          16\pi\Phi}
        g^{\mu \nu}\partial_\mu \Phi
        \partial_{\nu}\Phi
        \right ]\ \ . }
        \label{una}
        \end{equation}
Newton's constant $G=\Phi^{-1}$
or equivalently the dynamical Planck mass, $m_{pl}$, is
related to $\Phi$ by $m_{pl}=\Phi^{1/2}$.
The metric signature $(-,+,+,+)$ was used and  ${\cal R}$ is the scalar
curvature.
The action (\ref{una}) is completely general.
A given theory is specified by choosing the functional form
of $\om(\p)$.   It is explicitly enforced in (\ref{una})
that no potential nor cosmologogical constant is present.
Though the $\p{\cal R}$ coupling can be viewed as a potential
of sorts, variation of this term
with respect to
the metric leads to derivatives
on $\p$ (see eqn (\ref{enerphi})).
Therefore the $\p{\cal R}$ term can be identified
as contributing to the kinetic energy density.
A potential on the other hand would give a
contribution to $T_{\mu \nu}$ proportional to
$ g_{\mu \nu}$.

Varying the action with respect to $g_{\mu\nu}$ gives the
Einstein-like equations of motion
   \begin{equation}{
   {\cal R}_{\mu\nu}-{1\over2}g_{\mu\nu}{\cal R}
  = {8\pi\over\Phi}
    T^{\p}_{\mu\nu}\ \ .}
\label{likin}
   \end{equation}
The energy-momentum tensor for $\p$ is given by
     \begin{equation}
     T^{\p}_{\mu\nu}=
  {\om(\Phi)\over
8\pi\Phi}\left[\partial_\mu\Phi\partial_\nu\Phi-{1\over2}
  (\partial_\alpha\Phi \partial^\alpha\Phi)
  g_{\mu\nu}\right]
+{1\over 8\pi
  }\left(D_\mu
  D_\nu\Phi-g_{\mu\nu} \square\Phi\right )\ \
        \ \ ,
  \label{enerphi}
        \end{equation}
with $\square=g^{\mu\nu}D_{\mu}D_{\nu}$, and
$D_{\mu}$ is the covariant
derivative.  The first two terms in the energy momentum
tensor
(\ref{enerphi}) are analogous to those for a minimally
coupled scalar field.
However, unusual kinetic terms,
namely the last term two terms in eqn (\ref{enerphi}),
appear in the $\p$ stress tensor due to the variation of the
nonminimal $\Phi{\cal R}$ coupling in the
action.
These are the culprits which lead to
the behavior given particular attention in this
paper.

The variation of the action with respect to $\p$ gives
the equation of motion
        \begin{equation}
        -{2\om \over 3}\square \p+{\cal R}
        +{\partial_\mu \p\partial^\mu \p\over \p}
        \left [{\om \over \p}-{\partial \om \over
        \partial \p}\right ]=0
        \label{forgot}
        \end{equation}

Hereafter,  it is  assumed that the spatial gradients in $\p$ are negligible
so that $T^{\mu \nu}_\p$ is homogeneous and isotropic.
The
 Friedmann-Robertson-Walker (FRW) metric
reflects this symmetry; $g_{\mu \nu}=diag(-1,a^2,a^2,a^2)$
where $a$ is the scale factor.
The equation of
motion for $\p$ becomes
         \begin{equation}{
        {\ddot \Phi +3H\dot \Phi=
        -{1 \over (3+2\omega)}
        {d\omega\over d\Phi} \dot \Phi^2 }\ \ .}
        \label{one}\end{equation}
For economy of notation
define
        \begin{equation}
  {f(\p)\equiv (1+2\om(\Phi)/3)^{1/2}\ \ .}
    \label{def}
   \end{equation}
The $\p$ equation of motion has the immediate solution, of much use later,
   \begin{equation}
    {\dot \Phi={-C\over a^3 f}\ \ .}
    \label{first}
   \end{equation}
The constant of integration, $C$, can
be positive, negative, or zero.

 In an FRW metric,
the equation
of motion for the scale factor $a$ becomes
        \begin{equation}
        H^2+{\kappa \over a^2}
        =
        -{\dot\Phi\over\Phi} H +{\omega \over 6}\left({\dot \Phi
        \over \Phi}\right)^2   \; \ \ .
        \label{two}
        \end{equation}
Hereafter the metric is taken for simplicity to be flat, $\kappa=0$.
Solving eqn (\ref{two}) for $H$ yields
    \begin{equation}
   {H=-{\dot \Phi\over 2\Phi}(1\pm f) }
        \ \ .
  \label{useful}
   \end{equation}
There are two branches  for $H$ which can lead to some
confusion.  Notice if $f>1$, then the Hubble expansion
is positive if the upper sign is chosen for
$\dot \p>0$ and the lower sign is chosen for $\dot \p<0$.
If $f<1$, then $H<0$
only when $\dot \p>0$ (for
either branch) and $H>0$ only when $\dot \p<0$
(for either branch).

In an FRW cosmology, the energy density in the $\Phi$-field  is
       \begin{equation}
        {\rho_{\Phi}\equiv T_{0 \ \Phi}^{0}={\om\over 16\pi}
         {\dot \p ^2\over \p}-{3\over 8\pi}H\dot \p\ \ .}
         \label{rhoo}\end{equation}
Notice if eqn (\ref{useful}) is used in the expression for
$\rho_\Phi$ that
        \begin{equation}
        \rho_\Phi={3\over 32\pi}{\dot \Phi^2\over \Phi}
        \left ( f\pm 1\right )^2\ge 0
        \ \ .
        \label{good}
        \end{equation}
For a healthy theory, there ought to be no negative energy
or ghost modes.  Therefore the kinetic
energy density must be positive.
Classically at least, the kinetic energy density is
positive in an FRW metric as
long as $\om\ge -3/2$.
This can be verified by considering the conformally related
Einstein frame as is done in \S \ref{boo}.

The pressure
is $p_{\Phi}\delta^{i}_j=T^{i}_{j \ \Phi}$,
         \begin{equation}
             {p_{\p}=
         {\om \over 16 \pi}{\dot \p^2\over \p}
         -{1\over 8\pi}H\dot \p
       -{\dot \p^2\over 8\pi(3+2\om)}{\partial \om\over \partial \p}\ \  .}
        \label{press}\end{equation}
Notice the last term can be negative if $\partial \om/\partial \Phi>0$
\cite{un}.
This indicates that a changing value of $\om(\Phi)$ can lead to a
negative pressure in scalar-tensor theories, in the
absence of any potential.
If $\om$ is a positive constant, as was assumed in the
original Brans-Dicke model, the pressure is positive.
It can be verified, as argued in the introduction,
that $p_\p$ has a contribution from $-{dE_\p/dV}$
where $E_\p=\rho_\p V$ and $V=a^3$.
Thus a negative pressure signifies, in part, that the energy
in a unit volume grows as the universe expands.
As shown below, the changing structure
of $\om(\p)$ can in turn lead to an epoch of
accelerated cosmic expansion, as can a constant but
negative $\om$.

\subsection{Condition on $\om(\p)$}

The previous results can be exploited here to uncover
a condition
$\om(\p)$ must satisfy in order to drive the
scale factor to accelerate.
In general the acceleration of the scale factor is given by
    \begin{equation}
        {  {\ddot a\over a}=
        H^2+\dot H=-{4\pi\over 3\Phi}\left (\rho+3p\right )\ \ .}
        \label{accel}
        \end{equation}
While the weak energy principle requires generally that $\rho>0$,
the pressure can be negative.  [The
undecorated $\rho$ and $p$ will refer to unspecified energy
and pressure.]
In standard cosmology both
$\rho$ and $p$ are always positive so that the universe always
decelerates.  In the inflationary cosmology by contrast, $\rho=-p$ and
the universe accelerates with  $\ddot a/a=\Lambda$.
As derived above,
there can be a negative contribution to the
pressure from the Planck field due to the
changing structure of gravity.  A bound is given below
on the form of $\om(\Phi)$ which leaves the pressure
negative enough
to power an acceleration of the scale factor.

The results of eqns (\ref{rhoo}) and (\ref{press})
are put to use in eqn (\ref{accel}).
The acceleration is
           \begin{equation}
           {{\ddot a\over a}=
           -{\om \over 3}\left ({\dot \p\over \p}\right )^2
          +H\left ({\dot \p\over \p}\right )
          +{1\over 2}\left ({\dot \p\over\p}\right)^2{\p\over
          (3+2\om)}{\partial \om\over \partial \p}  \\ .}
           \label{accfull}   \end{equation}
Using (\ref{def}) and (\ref{useful}) in eqn (\ref{accfull}) gives
   \begin{equation}
   {{\ddot a\over a}=-{1\over 2}\left ({\dot \Phi\over \Phi}\right )^2 f
        \left [ f\pm1-{df\over d\ln
   \p }{1\over f^2}\right ]\\ .}
   \label{look}
     \end{equation}
In order  for $\ddot a>0$, the following condition must be
satisfied:
    \begin{equation}
    { f\pm1-{df\over d\ln
    \Phi }{1\over f^2}<0 \ \ .}
          \label{condition}
     \end{equation}
If the functional form of $f=(1+2\om(\Phi)/3)^{1/2}$ changes as the
universe evolves
such that it obeys the bound of (\ref{condition}), the scale
factor of the universe will accelerate.
It is easy to imagine that such a phase is entered and exited
smoothly.

There is one other possibility which can lead to
$\ddot a>0$.  The specific combination of a constant
but negative
$\om$ ($f<1$) with the minus branch in eqn (\ref{condition})
can give $\ddot a>0$.  As mentioned below eqn  (\ref{useful}),
this corresponds to an accelerated
{\it expansion} only if $\dot \p<0$.
The particular combination $\om=-1$,
$\dot \p<0$ with the minus branch, was studied
in reference \cite{Ram} in the context of
string theory.
In that reference, the authors considered the role
of the accelerated expansion in the graceful   exit
problem.  Perhaps if $\om(\p)$ is allowed
to vary, as opposed to being constrained to the value $-1$,
a graceful exit will be easier to execute.

Incidentally, if there is a contribution to the energy-momentum
tensor from ordinary matter of the perfect fluid form
$T^{\mu}_{\nu\ {\rm matter}}=diag(-\rho,p, p, p)$ and a potential
is included for the $\Phi$ field, then the full acceleration is given by
the lengthy expression
           \begin{eqnarray}
           {\ddot a\over a}=&-&{4\pi\over 3\p}
          \left [\rho+3p + {(\rho-3p)\over (1+2\om/3)}\right ]
           -{\om \over 3}\left ({\dot \p\over \p}\right )^2
          +H\left ({\dot \p\over \p}\right )
          +{1\over 2}\left ({\dot \p\over\p}\right)^2{\p\over
          (3+2\om)}{\partial \om\over \partial \p}  \\
        &+&{8\pi\over 3\p}\left [V -{1\over (1+2\om/3)}
        \left \{2V-\p{\partial V\over \partial \p}\right \}\right ]
        \ \ .
           \label{ugh}
    \end{eqnarray}
This puts a more complicated bound on $f$ which is undoubtedly
harder to meet.
This
possibility was discussed in references \cite{un}.

Three separate examples of gravity driven accelerations are considered
following
a comment on the causal physics relevant for inflation.

\subsection{ Inflation and the Horizon Problem}
\label{infla}

The horizon problem questions
how our entire observable universe could appear so homogeneous
and isotropic.
In standard cosmology such a large volume encompasses many
regions which were causally disconnected at early times.
The smoothness of the universe
across these regions appears to defy causal microphysics.
Inflation resolves this quandry by blowing up a region, causally
connected at early times, large enough to
envelop everything as far as the eye can see.

Before facing the demands of sufficient inflation,
a much weaker condition can be used to severely
restrict the range of $\om(\p)$ pertinent to inflation.
During an epoch of acceleration, constant comoving
scales cross outside of the Hubble
radius $H^{-1}$ only to cross back inside during a decelerating
phase.
The key to resolving the horizon problem is for the scales
which cross inside today to have been causally connected
before they crossed out.
Thus a nominal condition for the acceleration
to be relevant for inflation is simply that
        \begin{equation}
        d_\gamma>H^{-1}
        \ \ .
        \label{req}
        \end{equation}
The distance a photon travels, $d_\gamma$,
defines the extent of a causally connected region.
If this condition is
not met, then none of  the scales which cross outside $H^{-1}$ during
the acceleration are causally connected.
This is much weaker than a sufficient inflation
condition.
It is shown here that in the simplest case
of the single scalar model studied, the nominal condition
(\ref{req}) can be met only if $\om(\p)<0$
and if $\p$ drops.

First consider the equation of motion
(\ref{two}) rewritten as
        \begin{equation}
        \left (H+{\dot \Phi\over 2\Phi}\right )^2={1\over 4}f^2\left (
        {\dot \Phi\over \Phi}\right )^2
        .
        \label{alt}
        \end{equation}
Taking the square-root and reexpressing this equation gives
        \begin{equation}
        {d\ln(\Phi a^2)\over dt}=\pm  f{\dot \Phi\over \Phi}
        \; .
          \label{pm}
        \end{equation}
Assume for now that the upper sign holds for
$\p$ growing and the lower sign for
$\p$ decreasing.
Using (\ref{first}) and    integrating over $dt$ gives
        \begin{equation}
        \Phi a^2=| C|\int{dt'\over a'}
        \label{named}
                \end{equation}
up to a constant of integration.
Notice that from this we can deduce the particle horizon distance;
that is, the distance a photon has traveled since the beginning of time,
        \begin{equation}
        d_\gamma={\Phi a^3\over |C|}
        \; ,
        \end{equation}
up to a constant of integration.
For ease of comparison, $H$ can be written as
        \begin{equation}
        H={|C|\over (2a^3\Phi)}{(f\pm 1)\over f}
        \ \ .
        \end{equation}
Using the above two expressions
eqn (\ref{req})
becomes $f<\pm 1$.
Since $f=(1+2\om/3)^{1/2}$ is always positive,
the condition is impossible to meet
if $\p$ grows. If $\p$ drops, then (\ref{req})
demands
          \begin{equation}
        \om<0\ \ .
        \label{harsh}
        \end{equation}
According
to (\ref{good}) the classical energy density is positive
in an FRW metric
as long as $\om \ge -3/2$.
Thus $\om<0$ is not forbidden, at least not by the weak
energy condition.

Physically, constraint (\ref{harsh}) says the acceleration
is terribly weak.  From
eqn (\ref{look}) and (\ref{first})
it can be seen that $\ddot a$ is always
suppressed by $(a^3\p)^2$.
Roughly, if $\p$ grows then the acceleration
is weakened relative to the scenario if $\p $
drops.
This gives a feel for why $\p$ growing is prohibited entirely
but $\p$ dropping allows a narrow range of interest.
Only the third example delineated below will fall within this
range.

Not all possible branches were considered
above.
More generally, all possibilities can be summarized
by the condition
        \begin{equation}
        f<\pm \left ( {1-\delta\over 1+\delta}\right )
        \label{sever}
        \end{equation}
where
the constant of integration dropped from
(\ref{named}) is  included in $\delta={a_i^2\p_i/ a^2\p}$.
The subscript $i$ denotes initial values
and $\delta$ is always positive.
When $f>1$, the condition $H>0$ enforces the
branches chosen below eqn (\ref{pm}).
When $f<1$, the condition $H>0$ can only be met
when $\p$ drops.  Therefore the case of
$\p$ growing is totally excluded by the proceeding
arguments.  When $\p$ drops and $f<1$,
both branches are allowed in eqn (\ref{pm}).
For the plus branch, $\delta <1$
and for the minus branch, $\delta >1$.
The restriction (\ref{sever}) on $f$ is more severe.

\section{Example 1: Bouncing Universe}
\label{boo}

As a first example,
take $\Phi $ to grow,
so that the strength of gravity weakens as the universe evolves.
An $\om(\p)$, or equivalently $f(\p)$, which leads
to a universe with a bounce is chosen below.
The acceleration is not sufficient to be of interest
for inflation.  The interesting quality to this universe
is the nonsingularity.
Before proceeding with the details of the solution,
it is worth sketching
some properties of the resultant cosmology.
The evolution can
be traced back to an
 infinitely
large universe.  The space collapses down to finite size and
 then  bounces into an accelerated
expansion.  Eventually it settles down
into a decelerating phase.

If instead the universe begins at the moment
of the bounce, the evolution is the same but the
physical picture is different.
The universe begins nonsingular with
zero energy density.  The last property
has appeal since it costs nothing
to create a universe with zero energy, cold and empty.
The Hubble expansion is initially zero.  As the Planck mass
moves, the expansion accelerates.  Eventually the acceleration
ends and the expansion slows forever.

The general sketch outlined above can
be substantiated by solving
the equations of motion of the previous section
 with
    \begin{equation}
          {f(\p)={1\over
        \left [2 \ln(\bar \phi/\phi)\right ]^{1/2}}\ \ ,}
    \label{ex}
   \end{equation}
where $\bar \p$ is an arbitrary constant
and again $f=(1+2\om/3)^{1/2}$.
As already discussed positivity of the energy
density demands $\om\ge -3/2$ or $f\ge 0$.
This condition is automatically met here.
Example  (\ref{ex}) gives an acceleration of
     \begin{equation}
    { {\ddot a \over a}={1\over 2}
    \left ({\dot \Phi\over \Phi}\right )^2 f \ge 0\ \
      .}
   \label{see}
    \end{equation}
It is interesting to note that the form for $f$ in eqn (\ref{ex})
is the same  form recently considered in \cite{Damour}
to demonstrate that the presence of matter fields can attract
scalar-tensor theories toward general relativity.

In this example, $\Phi$ grows and the
strength of gravity weakens with time.
For $\dot \p>0$ consider the minus branch in eqn (\ref{useful})
for $H$
    \begin{equation}
   {H={\dot \Phi\over 2\Phi}(f-1) }
        \ \ .
   \label{useful2}
   \end{equation}
Eqn (\ref{useful2}) can be integrated
over $dt$ to find
    \begin{equation}
   {a=\bar a\left ({\bar \Phi\over \Phi}\right )^{1/2}\exp\left [
   {-\left({\ln \bar \Phi/ \Phi \over 2}\right )^{1/2}}\right ]\ \ ,}
    \label{solved}
   \end{equation}
where $\bar a$ is the arbitrary constant of integration.

The Planck field $\Phi$ and the scale factor
$a$ can be related to cosmic time parametrically.
Define a new variable $y$ from
   \begin{equation}
        \p=\bar \p \exp({-y^2})\ \ .
        \label{why}
        \end{equation}
In terms of $y$
the scale factor is
   \begin{equation}
   {a=\bar a\exp\left[{y^2\over 2}-{y\over \sqrt{2}}
   \right ]\ \ .}
  \label{inter}
   \end{equation}
Substituting this expression into the first integral of motion
(\ref{first}) and using $f=1/(\sqrt{2}y)$
gives the following integral
\begin{equation}
   {I\equiv -\int_{1/\sqrt{2}}^{y}\exp\left[{1\over 2}\left (y'-{3\over
\sqrt{2}}\right )^2
   \right ] dy'={e^{9/4}K\over \sqrt{2}}\Delta t\ \ ,}
  \label{integrate}
   \end{equation}
where
$\Delta t\equiv t-t_i$ and
$K=|C|/(\bar a^3\bar \p)$.  The limit of
integration is chosen so that $\Delta t=0$ at $\om=0$ ($f=1$).
Integrating $I$ gives the elapse of  cosmic time in terms of the
parameter $y$
     \begin{equation}
    {\Delta t=i{e^{-9/4}\sqrt{\pi}\over  K}\left
[erf(i(y/\sqrt{2}-3/2))-erf(i)\right ]
    \ \ .}
  \label{done}
   \end{equation}
If $y$ is allowed to run from $+\infty$ to $-\infty$, then time
runs from $-\infty$ to $+\infty$.

In figure 1, the scale factor and the Planck field $\Phi$ are shown
as functions of time.
In figure 2, the Hubble constant is drawn with time.
At time equals minus infinity,
$\Phi=0$ and $f=0$.  The scale factor was  infinite and
$H=-\infty$.
The universe, huge at minus infinity, contracts down to
finite size and then rebounds into an expanding phase.
The bounce occurs at $\Delta t=0$ with $\p=e^{-1/2}\bar \p$ and
$f=1$.

Alternatively
one could start the universe with $f=1$;
that is, impose the initial condition that the universe pop
into existence at $\Delta t=0$ with $f=1$ ($\om=0$).
Initially then $\Phi_i=e^{-1/2}\bar\Phi$ and
$a_i=\bar a e^{-1/4}$.  Notice also that
$\rho_{\Phi}=0$ and therefore $H=0$.
As time marches forward, $\Phi$ grows, $f$ grows, and the universe
begins to expand.  The universe had to accelerate to
go from $H=0$   to an
expanding cosmology.

At the value $\bar \p$, $\p$ hits a maximum  and then begins to
drop.  As it does so the acceleration ends
and the Hubble expansion slows.
The onset of deceleration is demonstrated in figure 2.
Eventually the universe expands to infinite size, the
expansion slows
asymptotically to zero.
As $\p$ drops toward zero, $f$ also approaches zero
and $\om$ approaches $ -3/2$.
The kinetic coupling $\om$ never falls below
$-3/2$ and therefore the energy density is always positive.

As a technical point, notice that
as $\Phi$ reaches its maximum $\bar \p$, $f=\infty$.  Although this
looks singular, the equations of motion remain regular.
To put this another
way, it is often said that the Planck mass is frozen and thus
scalar-tensor theories are driven to general
relativity in the limit of $\om\rightarrow \infty$.
More correctly, scalar-tensor
theories are driven to general relativity
in the lmit of $\om\rightarrow \infty\ \ $
{\it and} $\ \ \om^{-2}d\om/d\p\rightarrow 0$.
In terms of $f$ this means $f\rightarrow \infty$ and
$f^{-3}df/d\p\rightarrow 0$, as can be seen by studying the right
hand side of equation of motion (\ref{one});
                \begin{eqnarray}
        \ddot \p+3H\dot \p
        &=&-{1\over f}{df \over d\ln\p}{\dot \p^2\over
\p}\\
        &\propto &{1\over f^3}{df\over d \p}
        \ \ .
        \end{eqnarray}
Here, although $f\rightarrow \infty$, $f^{-3}df/d\p$ does not
approach zero.
Therefore the motion of $\p$ is not locked in when
$f\rightarrow \infty$.
As $\p$ reaches $\bar
\p$, $\dot \p$ passes through zero while at this moment
        \begin{equation}
        \ddot \p |_{\p=\bar \p}
        =-{K^2} \bar \p<0
        \ \ ,
        \end{equation}
where $K=|C|/ (\bar a^3\bar \p)$.
Since $\ddot \p<0$, $\p$ has hit a maximum at $\bar \p$
and then begins to decrease.
This is effected by $C \rightarrow -C$ in the equations of motion.

\section{Example 2:  Weak Burst of Acceleration}

More generally, a whole family of scalar-tensor
gravity models can be investigated
with forms for $f(\p)$  similar to the last example.
Consider the following
        \begin{equation}
        f=1+{2 b\over (\ln(\bar \p/\p))^{1/n}}
        \ \ .
        \label{many}
        \end{equation}
Again, $\bar \p$ is an arbitrary integration constant and
$f=(1+2\om/3)^{1/2}$.
Take $\p$ growing again and integrate the Hubble equation
as before
        \begin{equation}
        a=\bar a \exp\left[-{n\over n-1}b(\ln(\bar \p/\p))^{n-1\over n}
        \right ] \ \ .
        \end{equation}
The constant of integration is $\bar a$.
As an example which avoids the  terrain of
$\om<0$, let
$n=1/2$.  Then $f(\p)$ of (\ref{many}) is greater
than 1 for all values of $\p$.  The scale factor
becomes $a=\bar a\exp(b/\ln(\bar \p/\p))$.
If $\p=0$ initially, the universe begins with infinite Hubble
expansion, infinite energy density, and decelerates as it
evolves.

 Depending on $n$ and $b$ the theory specified by (\ref{many}) may
lead to an acceleration, for some range of $\p$.
For $b=50$ for example and $n=1/2$ an accelerating phase is entered
briefly, for a spirt,  and then deceleration begins again.
The acceleration occurs for values of
$\p$ roughly  between $0.24 \bar \p$ and $0.81 \bar \p$.

The Planck field $\p$ grows forever, asymptotically slowing to zero
as $\p\rightarrow \bar\p$ and
$f\rightarrow \infty$.  The scale factor grows infinitely
large and the Hubble expansion slows asymptotically to zero.

\section{Example 3:  Kinetic Inflation}
\label{kin}

The purpose of this third example is to offer an explicit
scalar-tensor theory which meets the nominal requirement
of \S \ref{infla}
that $d_\gamma >H^{-1}$.  Here it is demonstrated that if $\om$ is
allowed to drop below 0, but still greater than -3/2, then the scales
which cross
outside of $H^{-1}$ during the acceleration
are causally connected.  This is not intended as a model
of inflation.
In fact, inflation in this context may not succeed.
The question of inflation is given more complete
attention in a subsequent paper \cite{prep}.
A list of the conspicuous weaknesses in a realistic attempt at
successful inflation
can be compiled:
(1) Spatial gradients in the Planck field have explicitly been neglected
throughout.
This is equivalent to the assumption that the universe
begin homogeneous and isotropic.
Such an assumption is always made for inflation.  When
inflation is powered by a potential this supposition may be justified.
However, it is not clear that the assumption can be justified in
the case of
a kinetic inflation.
The acceleration is so weak it may be unable to dilute
inhomogeneities (or flatten the universe).
(2)   An obvious reheating mechanism, or rather heating mechanism,
does not reveal itself. Perhaps fluctuations in the
Planck field could be coaxed into creating
the hot soup of the early universe.
(3)  In the conformally related Einstein frame, the universe
appears to decelerate as shown in the next section.
However the condition of sufficient inflation
demands the Einstein frame scale factor accelerate
\cite{prep}.
Unless some unforseen subtleties
resolve this, successful completion of inflation looks unlikely.
More optimistically,
as the structure of gravity evolves it
can be arranged quite simply that
the acceleration ends, providing a long sought for
graceful exit.
Hopefully this possibility would not be so fine tuned
as to make
the exit graceless.

Regardless   of the application to the initial condition
problems of cosmology, it is still worthwhile to demonstrate
the scale crossings as such effects could be pertinent to   the
issue of density perturbations.
Consider the theory given by
        \begin{equation}
        f(\p)=\ln(\Phi/\bar \p)
         \ \ .
        \end{equation}
As in the proceeding examples, $\bar \p$ is an arbitrary constant
and $f=(1+2\om/3)^{1/2}$.
Integrating the Einstein-like eqn (\ref{two})
gives the scale factor as a function of $\p$,
        \begin{equation}
        a=\bar a
        \left({\bar \p\over \p}\right )^{1/2}
        \exp\left[-{1\over 4}(\ln(\p/\bar \p))^2\right ]
        \end{equation}
For concreteness take the Planck mass to begin infinite.
Initially the scale factor is zero and the universe is created singular
and decelerating.

Condition (\ref{condition}) states that the universe
accelerates if
        \begin{equation}
        1+\ln(\p/\bar\p)-{1\over \left (\ln(\p/\bar\p)\right )^2}<0
        \ \ .
        \label{lala}
        \end{equation}
Eqn (\ref{lala}) is only satisfied for values of $\p$
below roughly $e^1 \bar \p$,
or in terms of $\om$, roughly $\om < 0$.
Therefore
the universe decelerates from $\p=\infty$ down to $\p=e^1 \bar \p$.
Below that the universe accelerates for a burst.  In order to prevent
a true ghost from developing the behavior has to be shut
off artificially so that $\om$ does not drop below the
forbidden $-3/2$.

The point of this example is to consider
the causal structure of the universe.
Using the results of \S \ref{infla}
 the particle horizon,
$d_\gamma=\int_{t_i}^t {dt/ a}$, is
\begin{equation}
        d_\gamma={\p a^3\over |C|}
 \ \
        \end{equation}
and the Hubble radius is
        \begin{equation}
        H^{-1}={2\p a^3\over |C|}{f\over f+1}
        \ \ .
        \end{equation}
Comparing the two shows $d_\gamma >H^{-1}$
when $f$ drops below 1, or in other words
when $\om$ drops below 0.
In figure 3, the Hubble radius $H^{-1}$ is  drawn versus
the scale factor in a logarithmic plot.  Also shown are
an array of physical scales, constant in the comoving frame;
i.e. $\lambda=a \lambda_{constant}$.
During the early deceleration, $H^{-1}$ grows and
scales cross inside the Hubble radius.  When the era
of acceleration is entered, $H^{-1}$ drops.
As in the inflationary picture, constant comoving scales
cross outside the Hubble radius
while the expansion accelerates.
Since $d_\gamma$ exceeds $H^{-1}$, the scales
crossing outside $H^{-1}$ are indeed causally connected.

To reiterate, this example demonstrates explicitly a kinetic
acceleration which might be relevant for inflation.
As it was
argued in \S \ref{infla} would have to be the case,
$\om$ drops below zero as the Planck mass drops.

\section{Einstein Frame}
\label{conform}

The kinetic driven acceleration can be studied in the
conformally related
Einstein frame where the theory of gravity is the usual
Einstein theory with a fundamental Planck scale $M_o=10^{19}$ GeV.
[A study of inflation in the conformally related frame
will not be presented here but will be presented in
\cite{prep}.]
However
different the world appears, there is no experiment
observers in two conformally related frames could perform
which would distinguish one universe from a conformally
related counterpart.
All experiments involve the comparison of
scales, whether it be the length scale of a ruler
or the unit of time read on a clock.
In the Jordan-Brans-Dicke (JBD)
frame, it was implicitly assumed that ruler
lengths were constant and thus there was meaning to
the physical length units.
In other words, the JBD scale factor truly
contains complete information
about how distances, constant in comoving coordinates,
evolve in physical units.
If the rulers were constant in the JBD
frame then they would not be constant under the conformal
transformation.
Since in the Einstein frame rulers therefore change with time, the
physical length units change with time.  The scale factor
does
not contain  complete information.  Another scale factor must be
introduced which accounts for the changing
meaning of a physical length unit.
Once this is done,
the outcome of all experiments can be compared and
must be the same.

Perform a conformal transformation on the metric
        \begin{equation}
        g_{\mu \nu}=\Omega^2 \tilde g_{\mu \nu}
        \ \ ,
        \end{equation}
where $\Omega=M_o/\Phi^{1/2}$.
Under the conformal transformation the action becomes
        \begin{equation}
        A=\int d^4x\sqrt{-\tilde g}\left [
        {M_o^2\over 16 \pi}\tilde {\cal R} -{1\over 2}
        \tilde g^{\mu \nu}\partial_\mu \Psi\partial_\nu\Psi
        \right ]
        \end{equation}
The field $\Psi$ is tantamount to a rewriting
of the $\Phi$ field:
        \begin{equation}
        \Psi
        \equiv {M_o\over \sqrt{8\pi}}\int  {(\om +3/2)^{1/2}\over \Phi} d\Phi
        \ \ .
        \end{equation}
Notice that $\Psi$ is real and the energy
density in the $\Psi$-field, $\rho_\Psi=\dot \Psi^2/2$ is positive
as long as $\om\ge -3/2$.
The momentum associated with the field, $p_\Psi=\rho_\Psi$,
is always positive.

In addition to the conformal transformation, perform the coordinate
transformation
        \begin{eqnarray}
        d\tilde t&=&\Omega^{-1}dt \\
        \tilde a&=&\Omega^{-1} a
        \end{eqnarray}
so that the spacetime interval can be written in the usual FRW form,
        \begin{eqnarray}
        d\tilde s^2&=&\Omega^{-2}ds^2\\
        &=&\left [-
(\Omega^{-1}dt)^2+(\Omega^{-1}a)^2d\vec x^2 \right ]\\
        &=&\left [ - d\tilde t^2+\tilde a^2d\vec x^2
        \right ]
        \; . \label{transtom}
        \end{eqnarray}
The metric in $(\tilde t, \tilde a x^i)$ coordinates
is thus $\tilde g_{\mu \nu}=(-1, \tilde a^2,
\tilde a^2, \tilde a^2)$.
The evolution of the scale factor and of $\Psi$ can be found
directly in
the Einstein frame;
        \begin{equation}
        \tilde a\propto \Delta \tilde t^{1/3}
        \end{equation}
        \begin{equation}
        \tilde \Psi\propto \pm \ln(\Delta \tilde t)
        \end{equation}
where the upper sign refers to $\Psi$ growing and the lower
sign to $\Psi$ dropping.
Clearly the scale factor always decelerates.
The solution seems to know nothing about $\om(\p)$ and
thus does not distinguish
between different scalar-tensor theories until an observer is included.

 At first glance
it seems all information about the acceleration of   the
spacetime is lost in the Einstein picture where the
FRW universe is filled with an ordinary, minimally coupled scalar field.
There is no acceleration.
However, in order to properly pose and answer questions,
an observer, a test particle, must be included.
The test particle carries rulers and clocks with which to
make observations.

Include in the action a term for a test particle:
        \begin{equation}
        A_{\rm tester}=\int m\left [ -g_{\mu \nu}dx^\mu dx^\nu\right ]^{1/2}
        \; . \label{tester}
        \end{equation}
In the JBD frame, $m$ is the constant mass of the observer.
Under the conformal transformation the action becomes
        \begin{equation}
        A_{\rm tester}=\int \tilde m\left [-\tilde g_{\mu \nu}dx^\mu dx^\nu
        \right ]^{1/2}
        \; . \label{newmass}
        \end{equation}
where $\tilde m=\Omega m$.
This means that the length scale of our test particle is
        \begin{equation}
        \tilde \lambda=\Omega^{-1} \lambda
        \; , \label{scale}
        \end{equation}
and $\lambda=1/m$ is the constant wavelength of the observer
in the JBD frame.
Better said, the observer carries clocks and rulers.
The rulers, being constants in the original JBD frame,
have variable lengths in the Einstein frame.
The observer's rulers scale as
indicated by (\ref{scale}),
$\tilde L_{\rm ruler}={\Phi^{1/2}\over M_o} L_{\rm ruler}$.
As $\Phi$ evolves, so does the length
of the ruler.

Therefore
the scale factor $\tilde a$ does not tell all about physical
scales in the Einstein frame.
The true scale factor is not $\tilde a$ but in some sense,
$\tilde R$,
 the distance in ruler
units of two points at one unit of  comoving separation,
defined by
        \begin{equation}
        \tilde R={\tilde a \over \tilde L_{\rm ruler}}
        \  \ .
        \end{equation}
In terms of JBD quantities $\tilde R=R={a\over L_{\rm ruler}}$.
The distance in ruler units is a conformal invariant.
In fact,
all ratios of lengths should
be conformally invariant.
Both JBD and Einstein observers agree on the numerical
value of $\tilde R$.

The Einstein obsever can define the expansion parameter,
$\tilde H_R=\tilde R'/\tilde R$ where $'$ represents derivatives
with respect to $\tilde t$.  The parameter $\tilde H_R$
accounts not only for the expansion of the spacetime, but
also for the change in  the units of distance.
In the JBD frame, $\dot a=Ha$ grows during the gravity
driven acceleration.  In terms of Einstein variables,
$Ha=\tilde H_R\tilde a$.
The observer in the Einstein frame agrees this
quantity grows.
The JBD observer sees the
separation
between two points at fixed comoving distance
grow at an accelerated
rate due to the cosmic expansion.
Einstein observers interpret this acceleration as due to
the relative rate at which rulers change
and not just the expansion of spacetime.

As a last comment, the Einstein frame picture allows for a singularity in
the scale factor.  For the bouncing universe,
where there is no singularity in the JBD-frame,
this signals the break
down of the conformal factor; that is, the singularity is
in the conformal factor.
The Einstein time can be found as a function of the
parameter $y$ by integrating the coordinate transformation,
        \begin{equation}
        \Delta\tilde t=\left ({\bar \Phi^{1/2}\over   M_o}\right ) {2\over 3K}
        e^{-3y/\sqrt{2}}
        \ \ .
        \end{equation}
At $y=\infty$, $\Delta\tilde t=0$ and the
scale factor vanishes in the Einstein frame but is infinite in
the JBD frame.  The (inverse)
conformal transformation relating $\tilde a$
to $a$ through $\tilde a =\Omega^{-1} a$, diverges at $y=\infty$;
$\Omega^{-1}=\Phi^{1/2}/M_o\rightarrow \infty$.

\section{summary}

In a theory with a dynamical Planck mass the kinetic energy
density can have an associated negative pressure.
The pressure is negative when the kinetic coupling
factor, $\om$, grows with Planck mass according to the
bound defined in the paper.  The pressure
is also negative for a branch of the solutions
when the Planck mass drops and $\om$ is a negative constant.
By comparison,
the pressure of a free minimally coupled field
is always positive.

As a result of the unusual
negative pressure, the cosmic expansion can
accelerate.  The acceleration is unique as it
is driven solely by the kinetic energy in the Planck field.
No potential nor cosmological constant is present.

It is suggestive that scalar-tensor gravity alone
can accelerate the scale factor.  An
accelerated scale factor is one of the key ingredients used in
inflation.  This hints of a gravity driven or kinetic inflation.
Still, the acceleration by itself does not
lead to a successful inflationary model.
For a successful inflationary model,
the universe must inflate enough to
solve the cosmological horizon, flatness, and monopole problems.
There are indications this may be impossible.
For instance, it is worrisome that no acceleration is
apparent in the conformally related Einstein   frame.
This suggests the acceleration is not meaningful for inflation.
In this paper, it
was shown that the weakest minimal requirement that the acceleration
be relevant for inflation can only be satisfied in a theory
for which $\om$ drops   below zero as the Planck mass
drops.
Although this strongly
restricts the range of theories, the possibility is not ruled out
by the weak energy condition.
Positivity of
the energy density in an FRW universe demands only that
$\om\ge -3/2$.
Before an attempt is made at building a model of
kinetic inflation
a few questions must be addressed.  For one, it must be
shown that the universe remain smooth even after spatial
gradients are included.
Further, a viable heating mechanism is needed.
Lastly, the Einstein picture must be shown to be consistent.

For general illustration, examples were pursued which
demonstrated an era of accelerated expansion.
As an interesting consequence,
a nonsingular cosmology  with a bounce was found.
The universe could be created
nonsingular
at the moment of the bounce.
In this picture, the
cosmos begins cold and empty as
the energy density is zero.
The formation of the universe from nothing
would be energetically free, which suggests
a small barrier to spontaneous creation.
Although the creation of a
universe from nothing is an inherently quantum process, it is worth
noting that quantum gravity energy
scales are not entered
since the energy is zero while the Planck scale is finite.
As the Planck mass begins to roll, the
universe begins to expand.
This scenario offers a departure from an initial
big bang.  A transition from this cold and quiet
beginning to the fiery soup of the early universe
remains to be pursued.

\bigskip

\centerline{\bf Acknowledgements}

I extend special gratitude to Katie Freese for
labor and thought put into these and related ideas.
Thank you also  to Dick Bond, Neil Cornish, and
Glenn Starkman for their thoughts and time spent
on this project.
I am also grateful for the support of the
Jeffrey L. Bishop Fellowship.

\centerline{\bf Figure Captions}

Figure 1:
The scale factor and the Planck field $\p$ are
drawn as functions of time.  Notice the
scale factor begins infinite and contracts.
At $\Delta t=0$, the universe bounces into
an expanding phase.

Figure 2:
The Hubble expansion
$H$ grows and the universe accelerates as $\p$ grows
to $\bar \p$.  However, while $\p$ decreases
$H$ decreases and the universe enters a decelerating phase.

Figure 3:  The Hubble radius, $H^{-1}$ is  drawn versus
the scale factor in a logarithmic plot.  Also shown are
an array of physical scales, constant in the comoving frame;
i.e. $\lambda=a \lambda_{constant}$.
For a time, the Hubble radius increases indicating
the universe decelerates.
It reaches a maximum and the decreases while
the universe accelerates.
During deceleration, scales cross inside $H^{-1}$.
During acceleration, scales cross outside $H^{-1}$.

\end{document}